\documentclass[twocolumn,aps,prl,superscriptaddress]{revtex4}
\usepackage{graphicx}
\usepackage{amsmath}
\usepackage{bm}

\def\bea{\begin{eqnarray}}
\def\eea{\end{eqnarray}}
\def\be{\begin{equation}}
\def\ee{\end{equation}}

\begin{document}

\title{Nonadiabatic Effects in Ultracold Molecules via Anomalous Linear and Quadratic Zeeman Shifts}

\author{B. H. McGuyer}
\affiliation{Department of Physics, Columbia University, 538 West 120th Street, New York, NY 10027-5255, USA}
\author{C. B. Osborn}
\affiliation{Department of Physics, Columbia University, 538 West 120th Street, New York, NY 10027-5255, USA}
\author{M. McDonald}
\affiliation{Department of Physics, Columbia University, 538 West 120th Street, New York, NY 10027-5255, USA}
\author{G. Reinaudi}
\affiliation{Department of Physics, Columbia University, 538 West 120th Street, New York, NY 10027-5255, USA}
\author{W. Skomorowski}
\affiliation{Department of Chemistry, Quantum Chemistry Laboratory, University of Warsaw, Pasteura 1, 02-093 Warsaw, Poland}
\affiliation{Theoretische Physik, Universit\"{a}t Kassel, Heinrich Plett Stra{\ss}e 40, 34132 Kassel, Germany}
\author{R. Moszynski}
\affiliation{Department of Chemistry, Quantum Chemistry Laboratory, University of Warsaw, Pasteura 1, 02-093 Warsaw, Poland}
\author{T. Zelevinsky}
\email{tz@phys.columbia.edu}
\affiliation{Department of Physics, Columbia University, 538 West 120th Street, New York, NY 10027-5255, USA}

\begin{abstract}     
Anomalously large linear and quadratic Zeeman shifts are measured for weakly bound ultracold $^{88}$Sr$_2$ molecules near the intercombination-line asymptote.  Nonadiabatic Coriolis coupling and the nature of long-range molecular potentials explain how this effect arises and scales roughly cubically with the size of the molecule.  The linear shifts yield nonadiabatic mixing angles of the molecular states.  The quadratic shifts are sensitive to nearby opposite $f$-parity states and exhibit fourth-order corrections, providing a stringent test of a state-of-the-art \textit{ab initio} model.


\end{abstract}
\date{\today}
\maketitle

\newcommand{\w}{3.25in}

\newcommand{\Schematic}[1][\w]{
\begin{figure}[h]
\includegraphics*[width=3.5in]{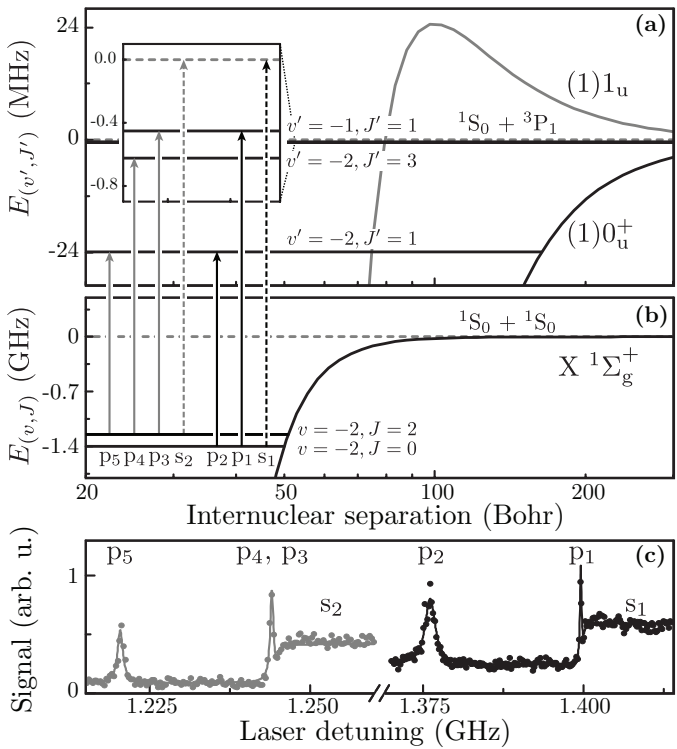}
\caption{$^{88}$Sr$_2$ potentials in the long range for (a) the excited state near the $^1S_0+{^3P_1}$ asymptote and (b) the ground state.  In the excited state, two least-bound vibrational levels with the total angular momentum $J'=1$ and $v'=-1,\;-2$ are shown.  As visible in the inset, the $|v',J'\rangle=|-2,3\rangle$ level is nearly degenerate with $|-1,1\rangle$.  In the ground state, the $v=-2$ molecules occupy the rotational levels $J=0,\;2$.  The allowed bound-bound transitions lead to spectral peaks and are labeled p$_1$ - p$_5$, while the bound-free transitions lead to 'shelf' lineshapes and are labeled s$_1$, s$_2$.  (c) A spectroscopic trace showing the five peaks and two continuum shelves is fitted with the expected lineshape.}
\label{fig:Sr2Schematic}
\end{figure}
}

\newcommand{\Zeeman}[1][\w]{
\begin{figure}[h]
\includegraphics*[width=3.5in]{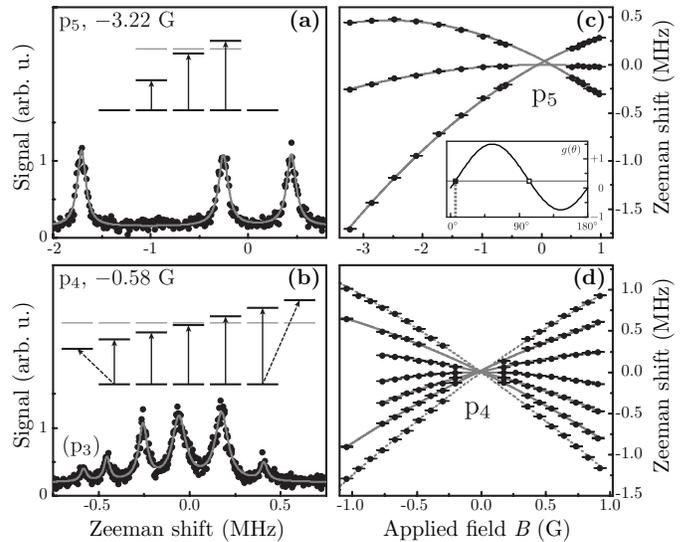}
\caption{Sample spectra of (a) p$_5$ and (b) p$_4$ in a small magnetic field $B$, showing $\pi$-transition peaks.  The p$_4$ trace includes a shifted sublevel of the nearby p$_3$.  (c, d) Their respective frequency shifts for varying field amplitudes are fitted to parabolic shapes with the required symmetry constraints.  In (d), the stretched levels with $|m'|=3$ are separately measured via $\sigma$-transitions.  The crossing point offset in (c) is likely caused by tensor light shifts from the optical lattice.  The inset in (c) shows the $g(\theta)$ curve from Eq. (\ref{eq:gTheta}), and indicates the measured Coriolis mixing angle $\theta=6.1^{\circ}$ for $|v',J'\rangle=|-2,1\rangle$.}
\label{fig:Sr2Zeeman}
\end{figure}
}

\newcommand{\MoleculeChi}[1][\w]{
\begin{figure}[h]
\includegraphics*[trim = 0in 0in 0in 0in, clip, width=3.5in]{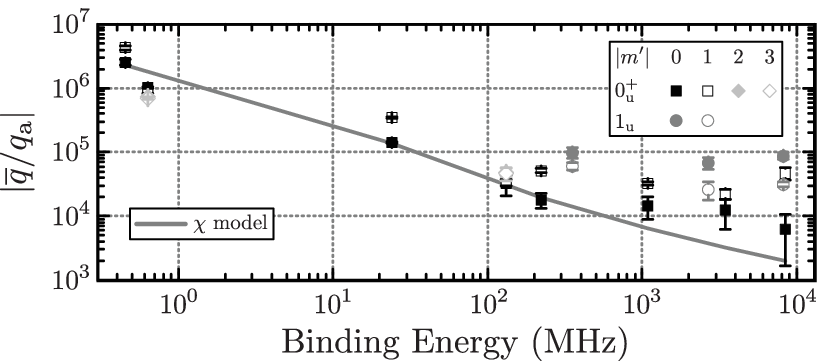}
\caption{Measured $\bar{q}$ versus the binding energy.  The values are normalized to the analogous atomic coefficient $q_a$, and show a million-fold enhancement for the weakest-bound levels.  The gray line represents the magnetic susceptibility model of Eq. (\ref{eq:ChiSimpleExpr}) for $J'=1$, $m'=0$ of $0_u^+$ that is in excellent agreement with the relevant data points (black squares for the $J'=1$ levels).}
\label{fig:Chi}
\end{figure}
}

Trapped ultracold molecules provide a rich testing ground for high-precision studies of quantum chemistry, few-body interactions, and many-body physics.  While ultracold atomic samples can be readily produced for many species, molecules present a far greater challenge due to the complexity of their rovibrational spectra.  Ground-breaking precision experiments can be performed with cold molecular beams or trapped molecular ions, such as measurements of the electron-to-proton mass ratio and its possible time variation, the electric dipole moment of the electron, and quantum electrodynamics in bound systems \cite{SchillerKoelemeijPRL07_HDIonPrecSpectrosc,ChardonnetShelkovnikovPRL08_muStability,HindsHudsonNature11_EDMYbF,ACMEArxiv13_ElectronEDM,UbachsDickensonPRL13_H2VibrPrecisMeast,PachuckiKomasaJCTC11_H2RovibQED}. However, ultracold temperatures allow extremely long coherence times and high spectral resolution for precise measurements in the optical domain.  Multiple methods have been developed to slow, cool, and trap simple molecules \cite{YeCarrNJP09_ColdMolecules}, including recent attempts by direct laser cooling \cite{DeMilleShumanNature10_SrFLaserCooling,YeHummonPRL13_YO_MOT} and evaporative cooling \cite{YeStuhlNature12_OHEvaporativeCooling}.  Nevertheless, currently the ultracold submicrokelvin temperature regime can be reached only by combining laser-cooled atoms into dimers.  Various alkali-metal atom pairs have been magnetoassociated near a Feshbach resonance to create high phase-space density samples \cite{NiSci08,LangPRL08_GroundStateRb2,NagerlDanzlNPhys10_Cs2GroundLattice}.  Recently, magnetic-field-insensitive alkaline-earth-metal atoms such as Sr were photoassociated into diatomic molecules \cite{ZelevinskyReinaudiPRL12_Sr2,SchreckStellmerPRL12_Sr2,TakahashiKatoPRA12_Yb2InLattice}, to take advantage of their spinless nature for many-body physics \cite{SchreckStellmerPRA13_SrRbQuantDegen} and precision measurements \cite{ZelevinskyPRL08}.

In this Letter, we report precise measurements and modeling of strong nonadiabatic effects in weakly bound ultracold $^{88}$Sr$_2$ molecules.  The molecules are trapped in an optical lattice and exposed to weak magnetic fields.  Nonadiabatic effects in this physical system involve Coriolis mixing of electronic and nuclear (rovibrational) dynamics.  These effects lead to strongly modified first-order (linear in magnetic field $B$) Zeeman shifts of the molecular energies near the $^1S_0+{^3P_1}$ intercombination atomic asymptote.  The linear shift coefficients yield the nonadiabatic mixing angles of the molecular wave functions.  Furthermore, anomalously large quadratic Zeeman shifts were observed, over a millionfold enhanced compared to those of free Sr atoms in the ${^3P_1}$ state.  Our \textit{ab initio} calculations for this heavy diatomic molecule agree on average to 5\% with the measurements of the linear shifts ($4\times$ the typical experimental uncertainty), and to $<20$\% with measurements of the quadratic shifts (matching the typical experimental uncertainty).  The quadratic shifts are shown to increase roughly cubically with the size of the molecule, by adopting a simple model of magnetic susceptibility (or magnetizability, the magnetic analogy of electric-dipole polarizability) that is enhanced by the proximity to the scattering continuum.  It is shown that the quadratic shifts are strongly affected by the opposite $f$-parity states, and that the most weakly bound levels experience substantial fourth-order Zeeman shifts even at $B\sim1$ G.  The Sr dimers are uniquely suitable for these measurements because the narrow intercombination transition (10 $\mu$s lifetime) allows a high spectral resolution, and because the very weakly bound states near this asymptote experience the strongest nonadiabatic effects.

\Schematic
Molecules of $^{88}$Sr$_2$ in their electronic ground state are created from a 1-$\mu$K gas of $^{88}$Sr via photoassociation near the 689 nm $^1S_0+{^3P_1}$ asymptote followed by spontaneous emission to a vibrational level with a large Franck-Condon overlap \cite{ZelevinskyReinaudiPRL12_Sr2}.  The resulting molecules are predominantly in the second vibrational level from the asymptote ($v=-2$), distributed between two rotational levels with total angular momenta $J=0,\;2$.  Figures \ref{fig:Sr2Schematic}(a) and \ref{fig:Sr2Schematic}(b) show the excited- and ground-state long-range $^{88}$Sr$_2$ potentials.
The ground state X dissociates to $^1S_0+{^1S_0}$, while the excited states $0_u^+$ and $1_u$ dissociate to $^1S_0+{^3P_1}$, where $|\Omega'|=0$ or $1$ in the state label refers to the total electronic angular momentum projection onto the internuclear axis \cite{HundsC}.
Primed labels refer to the excited electronic state, and negative vibrational indices count down from the asymptote.  Near the asymptote, ($\vec{L}\cdot\vec{J}$)-type Coriolis coupling strongly mixes the $0_u^+$ and $1_u$ potentials.  As a result, each excited level $|v',J'\rangle$ is a combination of orthonormal basis states, $|v'(|\Omega'|=0),J'\rangle$ and $|v'(|\Omega'|=1),J'\rangle$ .
A mixing angle $\theta$ determines each actually observed quantum state,
\be
|v',J'\rangle=\cos\theta|v'(0),J'\rangle+\sin\theta|v'(1),J'\rangle.
\label{eq:psi}
\ee
In the excited state, two least-bound vibrational levels with $J'=1$ and $v'=-1,\;-2$ are shown in Fig. \ref{fig:Sr2Schematic}(a).  The inset shows that $|v',J'\rangle=|-2,3\rangle$ is nearly degenerate with $|-1,1\rangle$.  (In this notation, the vibrational quantum numbers are assigned before the rotational ones.)  Note that spin statistics forbid odd values of $J$, as well as even values of $J'$ for $0_u^+$, while all $J'\geq1$ are allowed for $1_u$.  Figure \ref{fig:Sr2Schematic} shows the bound-bound transitions p$_1$ - p$_5$ (peaks) that are allowed by electric-dipole selection rules, and the bound-free transitions s$_1$ and s$_2$ (shelves).  A spectroscopic trace showing the five peaks and two shelves is displayed in Fig. \ref{fig:Sr2Schematic}(c).  Since the $v'=-1,\;-2$ levels both have large overlaps with the ground-state scattering continuum, these excited molecules spontaneously decay to the atomic ground state with a high probability, enabling a straightforward detection mechanism based on reappearance of Sr atoms in the lattice, after any nonphotoassociated atoms have been removed \cite{ZelevinskyReinaudiPRL12_Sr2}.
The probing of $^{88}$Sr$_2$ takes place in the Lamb-Dicke and resolved-sideband regimes of a one-dimensional magic-wavelength optical lattice \cite{LeibfriedRMP03,YeSci08} with parameters similar to those in Ref. \cite{ZelevinskyReinaudiPRL12_Sr2}.

\Zeeman
The spectra of $^{88}$Sr$_2$ molecules are studied under small static magnetic fields $|B|\lesssim3$ G.  The field coils are calibrated with Sr atoms in the lattice, assuming that the $g$ factor for the ${^3P_1}$ state is $g_a=1.5$.  The molecular spectra are obtained either directly by recovering the atoms with p$_1$ - p$_5$, or indirectly by using an additional spectroscopy laser that depletes $v=-2$ prior to recovering the atoms with p$_1$ - p$_5$, s$_1$, or s$_2$.  The former yields nearly background-free atom recovery peaks, and the latter results in dips corresponding to a depletion of the recovery signal.  While $\pi$ transitions were used for most measurements, the $g$-factor signs were determined with $\sigma$ transitions.  Sample spectra of p$_5$ and p$_4$ are shown in Figs. \ref{fig:Sr2Zeeman}(a) and \ref{fig:Sr2Zeeman}(b).  The p$_5$ spectrum is visibly asymmetric due to strong quadratic Zeeman shifts.  We confirmed that any Zeeman shifts of the ground-state sublevels are negligible for our precision, by systematically comparing p$_2$ and p$_5$.  Figures \ref{fig:Sr2Zeeman}(c) and \ref{fig:Sr2Zeeman}(d) show the frequency shifts plotted versus $B$ and fitted to parabolic shapes with the required symmetry constraints.
Large linear and quadratic shifts are apparent; no linear shifts are expected for these $0_u^+$ levels in the adiabatic picture.
The absolute values of linear Zeeman shifts were recently observed also in intercombination-line photoassociation experiments with Ca atoms \cite{RiehleKahmannArxiv13_CaPAZeeman}.

The Zeeman effect in the molecule arises from the interaction
\be
H_Z=\mu_B(g_L\vec{L}+g_S\vec{S})\cdot\vec{B},
\label{eq:HZ1}
\ee
where $\vec{L}$ and $\vec{S}$ are the electronic orbital and spin angular momenta, $g_L=1$, $g_S\approx2$, and $\mu_B$ is the Bohr magneton.
The resulting shift of each binding energy is
\begin{eqnarray}
\label{eq:qDefinedTh}
\Delta E_b&=g\mu_Bm'B+\sum\limits_{n>1}q_n\mu_BB^n \\
&\approx g\mu_Bm'B+\bar{q}\mu_BB^2,
\label{eq:qDefinedExp}
\end{eqnarray}
where $g\equiv\langle H_Z\rangle/(m'\mu_BB)$ depends on the expectation value of $H_Z$ for the given rovibrational level and the quantum number $m'$ is the projection of $J'$ onto the magnetic field axis.  Note that $q_n=q_n(v',J',m')$ depends on $m'$ while $g=g(v',J')$ does not, and that $\bar{q}$ incorporates residual effects of the higher even-order shifts.
The coefficients calculated from the \textit{ab initio} theory are based on Eq. (\ref{eq:qDefinedTh}), while the experimental data are fitted to Eq. (\ref{eq:qDefinedExp}) over the tested magnetic field range.

From Eqs. (\ref{eq:psi}) and (\ref{eq:HZ1}), the $g$ factor of the $|v',J'\rangle$ level is
\begin{equation}
g=g(\theta)=\frac{g_a\sin^2\theta}{J'(J'+1)}+\frac{g_a\sin2\theta}{\sqrt{J'(J'+1)}}\langle v'(0)|v'(1)\rangle,
\label{eq:gTheta}
\end{equation}
where the positive overlap of the vibrational parts of the basis states is $\langle v'(0)|v'(1)\rangle\sim1$.  Theoretical mixing angles $\theta$ were obtained from the updated \textit{ab initio} model of Ref. \cite{MoszynskiSkomorowskiJCP12_Sr2Dynamics}.  The model couples seven electronic potentials with constraints from data in Ref. \cite{TiemannSteinEPJD10_Sr2XPotential} and reproduces most of the measured $J'=1,3$ resonances at the percent-level accuracy through the tuning of long-range parameters and couplings in the Hamiltonian in Eq. (20) of Ref. \cite{MoszynskiSkomorowskiJCP12_Sr2Dynamics}.  Pure $0_u^+$ levels ($\theta=0$) do not exhibit linear Zeeman shifts; pure $1_u$ levels ($\theta=90^{\circ}$) have $g=0.75$ for $J'=1$.  The inset in Fig. \ref{fig:Sr2Zeeman}(c) shows $g(\theta)$ from Eq. (\ref{eq:gTheta}) for $J'=1$, assuming a vibrational overlap of 1 (this is accurate to $\sim10\%$ for the measured states).  The Coriolis mixing angle $\theta$ varies from 0 to $\pi$ and thus includes both the relative amplitude and the phase of the two wave function components.  For each measured $g$-factor there are two solutions for $\theta$, as indicated in the inset.  The experimentally determined angle pair for $|v',J'\rangle=|-2,1\rangle$ in the inset is $\theta=\{6.1^{\circ},103.4^{\circ}\}$, the first value closely agreeing with the \textit{ab initio} angle $\theta=5.9^{\circ}$.

The quadratic Zeeman shifts due to $H_Z$ can be calculated for each level by applying standard second-order perturbation theory and summing the contributions from the continuum and bound rovibrational levels that are coupled by $H_Z$ with the given level.  The first group of contributing levels includes those with the same $J'$ (1 or 3) that belong to the coupled $0_u^+$ and $1_u$ potentials; this contribution exists only for $m'\neq0$ and is negative for $0_u^+$ levels and positive for $1_u$ levels.  The second group includes levels with even $J'$ (2 or 4) and opposite $f$ parity that belong to the $1_u$ manifold; this contribution is negative and significant for all $m'$.  For the four most weakly bound levels, the dominant contribution to the quadratic shift is from the continuum of scattering states above the $^1S_0+{^3P_1}$ asymptote, while for the deeper levels it is from the nearest bound level.  The correct prediction of the quadratic Zeeman shifts requires an accurate description of the continuum and bound levels with $\Delta J'=0,\pm1$; thus, these shifts provide a substantially more stringent test of the molecular model than the linear shifts alone.  Any inaccuracy of the electronic potentials in the \textit{ab initio} model affects nonadiabatic mixing and thus the Zeeman shifts more strongly than the binding energies; moreover, for weakly bound levels small errors in binding energies can lead to significant errors in second-order properties.

Table \ref{tab:Values} presents the measurements and \textit{ab initio} calculations of the quadratic coefficients $\bar{q}$ from Eq. (\ref{eq:qDefinedExp}), showing an average disagreement roughly matching the experimental uncertainty.
We find that for the most weakly bound level, the experiment is sensitive to fourth-order Zeeman shifts, and these are included in the quoted $\bar{q}$ coefficients.
The measured and calculated $g$-factors are also shown; the average agreement is $5\%$ if the two most deeply bound levels with strong mixing are excluded, while the typical experimental uncertainty just exceeds $1\%$.  The linear Zeeman shift measurements allow a (model-dependent) experimental determination of the Coriolis mixing angle from Eq. (\ref{eq:gTheta}), shown in the last column.  These angles agree with \textit{ab initio} calculations to a few percent.  For the deepest level pair the disagreement with theory was large enough that we were unable to determine which of the two possible angles is correct.
\begingroup
\squeezetable
\begin{table}[t!]
\caption{\label{tab:Values} Magnetic properties relevant to nonadiabatic effects in weakly bound levels of $^{88}$Sr$_2$ near the $^1S_0+{^3P_1}$ asymptote, ordered by increasing binding energy $E_b$ (MHz). The effective experimental and theoretical quadratic shift coefficients $\bar{q}$ (G$^{-1}$) are defined in Eq. (\ref{eq:qDefinedExp}).  The starred values indicate strong fourth-order contributions [$q_4$ in Eq. (\ref{eq:qDefinedTh})].  The experimental and theoretical $g$ factors are also shown, including their signs.  The (model-dependent) measured Coriolis mixing angles $\theta$ are listed; two angles are possible for each of the two deepest levels due to the limitations of the model.}
\begin{ruledtabular}
\begin{tabular}{l c c c c c l l}
$E_\text{b}$    & $J'$  & $g$        & $g$ & $\theta$           & $|m'|$        & $-\bar{q}$       & $-\bar{q}$        \\
      &               &         (Expt.)                      &      (Theory)                 &          (Expt.)                             &               & (Expt.)            & (Theory)    \\
\hline
0.45                    & 1             & 0.666(14)             & 0.636                 & 16.5$^{\circ}$                & 0             & 0.325(34)$^*$             & 0.266$^*$  \\
                        &               &                               &                       &                                       & 1             & 0.546(44)$^*$             & 0.780$^*$  \\
0.63                    & 3             & 0.270(2)              & 0.271         & 18.5$^{\circ}$                & 0             & 0.130(5)              & 0.114         \\
                        &               &                               &                       &                                       & 1             & 0.102(5)              & 0.113         \\
                        &               &                               &                       &                                       & 2             & 0.095(6)              & 0.112          \\
                        &               &                               &                       &                                       & 3             & 0.090(4)              & 0.110          \\
24.0                    & 1             & 0.232(3)              & 0.222         & 6.1$^{\circ}$                         & 0             & 0.0181(6)             & 0.0147         \\
                        &               &                               &                       &                                       & 1             & 0.0444(13)            & 0.0388 \\
132                     & 3             & 0.173(2)              & 0.160         & 11.6$^{\circ}$                & 0             & 0.0041(15)            & 0.0039 \\
                        &               &                               &                       &                                       & 1             & 0.0043(5)             & 0.0042        \\
                        &               &                               &                       &                                       & 2             & 0.0061(9)             & 0.0050        \\
                        &               &                               &                       &                                       & 3             & 0.0060(14)            & 0.0062        \\
222.2           & 1             & 0.161(2)              & 0.148         & 4.2$^{\circ}$                         & 0             & 0.0023(6)             & 0.0022        \\
                        &               &                               &                       &                                       & 1             & 0.0066(4)             & 0.0057 \\
353.2           & 1             & 0.625(9)              & 0.610         & 93.3$^{\circ}$                & 0             & 0.0126(26)            & 0.0111  \\
                        &               &                               &                       &                                       & 1             & 0.0077(10)            & 0.0065        \\
1084.1          & 1             & 0.142(2)              & 0.128         & 3.8$^{\circ}$                         & 0             & 0.0019(7)             & 0.0013        \\
                        &               &                               &                       &                                       & 1             & 0.0041(3)             & 0.0031 \\
2683.7          & 1             & 0.584(8)              & 0.571         & 94.6$^{\circ}$                & 0             & 0.0080(18)            & 0.0067  \\
                        &               &                               &                       &                                       & 1             & 0.0034(11)            & 0.0031  \\
3463.3          & 1             & 0.193(3)              & 0.174         & 5.1$^{\circ}$                         & 0             & 0.0016(8)            & 0.0017 \\
                        &               &                               &                       &                                       & 1             & 0.0029(5)             & 0.0028  \\
8200.2          & 1             & $-$0.149(2)           & $-$0.592      & \{113.6$^{\circ}$,            & 0             & 0.0112(12)            & 0.0076 \\
                        &               &                               &                       & 175.9$^{\circ}$\}             & 1             & 0.0040(4)             & 0.0017 \\
8429.7          & 1             & 0.931(13)             & 1.333         & \{24.6$^{\circ}$,             & 0             & 0.0008(6)             & 0.0009 \\
                        &               &                               &                       & 84.9$^{\circ}$\}              & 1             & 0.0060(13)            & 0.0047
\end{tabular}
\end{ruledtabular}
\end{table}
\endgroup

\MoleculeChi
The $\bar{q}$ measurements are plotted in Fig. \ref{fig:Chi} versus the binding energy; they are normalized to the estimated analogous coefficient of ${^3P_1}$ Sr atoms, $q_a=1.28\times10^{-7}$/G.
The anomalously large quadratic Zeeman shifts grow by several orders of magnitude with decreasing binding energy.  This behavior can be qualitatively understood by considering the magnetic susceptibility $\chi$, such that the second-order shifts are $\Delta E_b^{(2)}=-\chi B^2/2$.  The susceptibility for $0_u^+$ is
\be
\chi(R)\approx C_{J'}^{m'}\frac{\mu_B^2}{V_1(R)-V_0(R)}\approx C_{J'}^{m'}\frac{\mu_B^2R^3}{3C_3},
\label{eq:ChiSimpleExpr}
\ee
where $R$ is the internuclear separation, $V_{|\Omega'|}$ denote the potentials in Fig. \ref{fig:Sr2Schematic}(a), and the coefficient $C_{J'}^{m'}$ resulting from transformation of the magnetic susceptibility tensor from the molecule-fixed to the space-fixed frame \cite{BrownCarrington} is of order unity.  Equation (\ref{eq:ChiSimpleExpr}) assumes that near the asymptote, $C_3$ terms dominate the long-range potentials.
This gives the paramagnetic component of the susceptibility; the diamagnetic component is negligible.
The line in Fig. \ref{fig:Chi} shows $\chi$ for the $J'=1$, $m'=0$ levels of $0_u^+$, with $C_3$ taken from Ref. \cite{ZelevinskyPRL06} and $R$ set to the classical outer turning points.  The $\chi$ model shows an excellent agreement with the relevant data points (black squares on the plot for $J'=1$).  This model is more reliable for the levels where the largest contributions to $\chi$ come from the continuum, justifying approximating the energy difference as $V_1(R)-V_0(R)$.

The large enhancement of quadratic Zeeman shifts near the atomic asymptote highlights the fact that two bound atoms separated by hundreds of Bohr radii have very distinct properties from a pair of free atoms.  To explore this further, we slightly changed the molecular potentials to move the least-bound level closer to the threshold.  The cubic scaling with $R$ is preserved for the smallest binding energies allowed by the $J'=1$ centrifugal barrier, leading to quadratic Zeeman shifts about $10\times$ greater than reported here.  As evident from Eq. (\ref{eq:gTheta}), even the linear Zeeman shifts cannot be directly related to the atomic values.  The difference between molecular and atomic properties is more pronounced for the quadratic effect, since it results from coupling to all possible molecular states, both bound and continuum.  We expect that any features of the continuum near the threshold, such as shape or Feshbach resonances, would be also strongly modified by magnetic fields.

In conclusion, we have presented precise measurements of strongly enhanced nonadiabatic effects in ultracold molecules.  This work was carried out through experimental and \textit{ab initio} studies of magnetic Zeeman shifts in weakly bound $^{88}$Sr$_2$ near the $^1S_0+{^3P_1}$ intercombination-line asymptote.  The molecules were produced and optically probed in the tight-confinement regime of a magic-wavelength optical lattice.  For a series of bound levels, molecular $g$ factors were measured and calculated with good agreement.  Furthermore, quadratic shifts were observed and accurately modeled both with the full \textit{ab initio} approach and with an approximate magnetic-susceptibility model.  The $g$-factor measurements yield accurate nonadiabatic mixing angles of the molecular wave functions.  $^{88}$Sr$_2$ presents unusual opportunities for high-precision studies of molecular physics due to very weakly bound levels that can be optically resolved, and because of accurate state-of-the-art \textit{ab initio} molecular modeling.  These measurements of nonadiabatic mixing between molecular states, together with the \textit{ab initio} model, constitute one of the most precise tests of modern quantum chemistry.

This work was partially supported by the ARO grant W911NF-09-1-0504 and the Sloan Foundation.
R. M. thanks the Polish Ministry of Science and Higher Education for support through the project
N-N204-215539 and the Foundation for Polish
Science for support within the MISTRZ program.  M. M. acknowledges the NSF IGERT (DGE-1069260), and W. S. the Alexander von Humboldt Foundation, for support.  We are grateful to G. Iwata for contributions to the experiment.


\end{document}